\documentclass{article}
\usepackage{amssymb}

\input{tcilatex}

\begin{document}

\textbf{RANDOM SHOOTING OF ENTANGLED PARTICLES IN VACUUM}

\bigskip

E. A. Novikov

\bigskip

Institute for Nonlinear Science, University of California - San Diego, La
Jolla, CA 92092-0402

\bigskip

The effect of random shooting of particles is considered on the basis of
solution of the Schr\"{o}dinger equation and in terms of the Wigner
function. Two-particle description shows, in particular, that initial
correlation leads to high velocities of particles. This could be a potential
mechanism for obtaining energy. Evolution of the n-particle probability
distribution is described analytically.

\bigskip

\bigskip

The Heisenberg's uncertainty relation reads:

\begin{equation}
\Delta p\Delta x\sim \text{%
%TCIMACRO{\U{127}}%
%BeginExpansion
h{\hskip-.2em}\llap{\protect\rule[1.1ex]{.325em}{.1ex}}{\hskip.2em}%
%EndExpansion
}  \tag{1}
\end{equation}%
where $\Delta p$ is uncertainty of momentum ( $p=mv$ - product of mass and
velocity of particle), $\Delta x$ is uncertainty of coordinate and 
%TCIMACRO{\U{127} }%
%BeginExpansion
h{\hskip-.2em}\llap{\protect\rule[1.1ex]{.325em}{.1ex}}{\hskip.2em}
%EndExpansion
is Planck`s constant. Relation (1) seems to preclude the existence of
trajectory: the more exactly we know location the more uncertain is
velocity. Probability density of particle position is given by $P=\Psi \Psi
^{\ast }$ , where $\Psi (t,x)$ is the wave function and star indicates
complex conjugate. For simplicity, we start with the one-dimensional case,
but the 3-D generalization is simple (see below).

Schr\"{o}dinger`s equation has the form (see, for example, Ref. 1):

\begin{equation}
i\text{%
%TCIMACRO{\U{127}}%
%BeginExpansion
h{\hskip-.2em}\llap{\protect\rule[1.1ex]{.325em}{.1ex}}{\hskip.2em}%
%EndExpansion
}\partial \Psi /\partial t=-(\text{%
%TCIMACRO{\U{127}}%
%BeginExpansion
h{\hskip-.2em}\llap{\protect\rule[1.1ex]{.325em}{.1ex}}{\hskip.2em}%
%EndExpansion
}^{2}/2m)\partial ^{2}\Psi /\partial x^{2}+U(x)\Psi  \tag{2}
\end{equation}

Here $U(x)$ is the potential energy of the particle in the external field, $%
i $ is the imaginary unit. The strange behavior of particles is not
associated with particular form of potential energy. So, we put $U=0$ and
consider the influence of the quantum vacuum on the motion of particle.
Equation (2) becomes very simple:

\begin{equation}
\partial \Psi /\partial t=iq\partial ^{2}\Psi /\partial x^{2},\;q\equiv 
\text{%
%TCIMACRO{\U{127}}%
%BeginExpansion
h{\hskip-.2em}\llap{\protect\rule[1.1ex]{.325em}{.1ex}}{\hskip.2em}%
%EndExpansion
/}2m  \tag{3}
\end{equation}%
where $q$ is the coefficient of imaginary diffusion (ID). Equation for $\Psi
^{\ast }$ has the same form as (3) with minus in the right hand side.

General solution of equation (3) in unbounded space can be obtain with use
of Fourier transform:

\begin{equation}
\partial \widetilde{\Psi }/\partial t=-iqk^{2}\widetilde{\Psi }  \tag{4}
\end{equation}%
where $k$ is the wave number and $\widetilde{\Psi }(t,k)$ is the transform
of $\Psi (t,x)$. Solution of (4) is:

\begin{equation}
\widetilde{\Psi }(t,k)=\widetilde{\Psi }_{0}(k)\exp \{-iqk^{2}t\}  \tag{5}
\end{equation}%
where zero indicates initial function. The inverse transform of (5) gives:

\begin{equation}
\Psi (t,x)=(4\pi iqt)^{-1/2}\int\limits_{-\infty }^{\infty }dx^{\prime }\Psi
_{0}(x^{\prime })\exp \{-\frac{(x-x^{\prime })^{2}}{4iqt}\}  \tag{6}
\end{equation}

We would like to clarify the physical sense of ID and expressions (5) and
(6).

Let us introduce Gaussian random velocity $v(t)$ with zero mean ($<v(t)>=0$)
and correlation proportional to $\delta $-function (see, for example,
textbook [2]):

\begin{equation}
<v(t+\tau )v(t)>=q\delta (\tau )  \tag{7}
\end{equation}

Here and below brackets $<>$ indicate statistical averaging. Consider
trajectory, produced by such velocity:

\begin{equation}
z(t)=z_{0}+\int\limits_{0}^{t}v(t^{\prime })dt^{\prime }  \tag{8}
\end{equation}

We assume that initial position $z_{0}$ has Gaussian distribution,
independent of $v(t)$, with zero mean ($<z_{0}>=0$) and dispersion $%
<(z_{0})^{2}>=a^{2}$. Then, $z(t)$ will have Gaussian distribution with zero
mean and dispersion:%
\begin{equation}
<[z(t)]^{2}>=a^{2}+qt  \tag{9}
\end{equation}

Here we used (7) and independence of $v(t)$ from $z_{0}$.

Now, consider complex trajectory:

\begin{equation}
y(t)=y_{0}+(1+i)\int\limits_{0}^{t}v(t^{\prime })dt^{\prime }  \tag{10}
\end{equation}

Here $y_{0}$ is real with the same statistics as $z_{0}$. We have $<y(t)>=0$
and

\begin{equation}
<[y(t)]^{2}>=a^{2}+2iqt  \tag{11}
\end{equation}

High order statistical moments of $y(t)$ can be calculated from these first
two by the same formulas as for the Gaussian distribution. Effective
probability density (epd) for $y_{1}=y-<y>$%
\[
(2\pi <y_{1}^{2}>)^{-1/2}\exp \{-\frac{y_{1}^{2}}{2<y_{1}^{2}>}\} 
\]%
can be used, as if $y$ is real. For statistical moments of $y^{\ast }(t)$ in
all formulas we have to change the sign in front of $i$. We will call such
complex random processes complex-Gaussian.

Returning to (6), for simplicity of calculations, let us choose:

\begin{equation}
\Psi _{0}(x)=(2\pi a^{2})^{-1/4}\exp \{-x^{2}/4a^{2}\}  \tag{12}
\end{equation}

This wave function corresponds to the Gaussian probability density for the
initial position of particle:

\begin{equation}
P_{0}(x)=(2\pi a^{2})^{-1/2}\exp \{-x^{2}/2a^{2})  \tag{13}
\end{equation}

Substitution of (12) into (6), after simple calculation, gives:

\begin{equation}
\Psi (t,x)=\frac{a^{1/2}}{(2\pi )^{1/4}(a^{2}+iqt)^{1/2}}\exp \{-\frac{x^{2}%
}{4(a^{2}+iqt)}\}  \tag{14}
\end{equation}

This expression (apart from normalizing factor) is the epd for the
complex-Gaussian process like (10) with $a$ replaced by $\sqrt{2}a$. It
suggests that quantum particles have hidden complex trajectories. It also
suggests that the quantum vacuum interacts with particles by producing
tachyonic impulses (with imaginary energy), which leads to ID (see below).

Using (14), we calculate the probability density for the position of
particle in real world:

\begin{equation}
P(t,x)=\Psi \Psi ^{\ast }=\frac{1}{[2\pi (a^{2}+w^{2}t^{2})]^{1/2}}\exp \{-%
\frac{x^{2}}{2(a^{2}+w^{2}t^{2})}\}  \tag{15}
\end{equation}

Here $w=qa^{-1}$ is effective constant velocity. The trajectory for (15) is
random shooting:

\begin{equation}
x(t)=x_{0}+ut  \tag{16}
\end{equation}

where $x_{0}$ and $u$ are independent random constants with Gaussian
distributions and $<x_{0}>=0,\;<(x_{0})^{2}>=a^{2},\;<u>=0,\;<u^{2}>=w^{2}$.
The fact that $w$ is inversely proportional to $a$ is, of course,
manifestation of the uncertainty relation (1).

Note that if we multiply the initial wave function (12) by $\exp \{i\gamma
x\}$, where $\gamma $ is real constant, the initial probability density (13)
will not change. However in (15) $x$ will be replaced by $x-2q\gamma t$, so
we will have $<x>=2q\gamma t$. In this paper we choose initial conditions
such that the mean positions of particles will not change in time.

The quadratic time-dependence of dispersion in (15) shows that quantum
complex trajectories like (10) can not be just projected on real axis,
leading to linear time-dependence (9). The nonlinear operation $\Psi \Psi
^{\ast }$ in (15) is proportional to the joint epd for the complex and
complex conjugate trajectories, as if they are independent. This indicates
an interesting sort of interaction. It is remarkable that such interaction
produces random shooting (16). Note that the effective energy $mw^{2}/2$ is
inversely proportional to $m$.

Let us look into details of this interaction by considering the
characteristic function:%
\begin{equation}
\Phi (t,k)=\int dxP(t,x)\exp \{ikx\}=\int dx\Psi (t,x)\Psi ^{\ast }(t,x)\exp
\{ikx\}  \tag{17}
\end{equation}

Using Fourier transformation, we get from (17):

\begin{equation}
\Phi (t,k)=\frac{1}{2\pi }\int dm\widetilde{\Psi }(t,m)\widetilde{\Psi }%
^{\ast }(t,m-k)  \tag{18}
\end{equation}

The transform of (12) is:

\begin{equation}
\widetilde{\Psi }_{0}(k)=2^{3/4}\pi ^{1/4}a^{1/2}\exp \{-a^{2}k^{2}\} 
\tag{19}
\end{equation}

Solution (5) gives:%
\begin{equation}
\widetilde{\Psi }(t,k)=2^{3/4}\pi ^{1/4}a^{1/2}\exp \{-(a^{2}+iqt)k^{2}\} 
\tag{20}
\end{equation}

From (18) we now have:

\begin{equation}
\Phi (t,k)=2^{1/2}\pi ^{-1/2}a\int dm\exp
\{-a^{2}[m^{2}+(m-k)^{2}]-iqt[m^{2}-(m-k)^{2}]\}  \tag{21}
\end{equation}

For $t=0$ :

\begin{equation}
\Phi (0,k)=\exp \{-\frac{1}{2}a^{2}k^{2}\}  \tag{22}
\end{equation}

which corresponds to the transform of (13).

Let us stress the difference between the ordinary diffusion, when $iq\equiv
\lambda $ is real, and our case of ID. For real $\lambda $ in the second
square bracket in (21) we will have sign plus between $m^{2}$ and $(m-k)^{2}$
- the same as in the first square bracket. For real $\lambda $ we will have
solution of the original equation (3) for the probability density (instead
of wave function or epd) with the linear time-dependence of the dispersion.
In our case of ID the sign inside the second square bracket is minus. This
leads to elimination of the linear time-dependence and to the solution:%
\begin{equation}
\Phi (t,k)=\exp \{-\frac{1}{2}(a^{2}+w^{2}t^{2})k^{2}\}  \tag{23}
\end{equation}

which is the transform of (15). The elimination effect can be considered as
interaction between $m$-tachyon and $(m-k)$-antitachyon. This interaction we
see after averaging. It will be very interesting to investigate in future -
what kind of "battle of tachyons" takes place before averaging.

There is another way of looking at the random shooting of particle with the
characteristic velocity $w=qa^{-1}$. Consider a pair of vortices with
intensities $q$ and $-q$, placed on the line perpendicular to real line in
the complex plane: one above the real line, another below on the same
distance $\sim a$ . These vortices will move each other and the particle
with velocity $\sim w$. This picture can be considered as coherent
structure, produced by the flow of tachyons.

The 3-D generalization of ID is natural to do in terms of characteristic
function with normalization condition $\Phi (t,0)=1$. For simplicity, we
assume isotropy of initial probability. Substitution of 3-D Laplacian $%
\Delta =\partial ^{2}/\partial x_{1}^{2}$ $+\partial ^{2}/\partial
x_{2}^{2}+\partial ^{2}/\partial x_{3}^{2}$ instead of $\partial
^{2}/\partial x^{2}$ in equation (3) will not change the isotropy. It means
that in expressions (12) - (15) we have now $%
x^{2}=x_{1}^{2}+x_{2}^{2}+x_{3}^{2}$ and prefactors are rased to the third
power. In equation (23) we now have $k^{2}=k_{1}^{2}+k_{2}^{2}+k_{3}^{2}$.
Correspondingly: $<x_{b}>=0,(b=1,2,3),\;<x_{b}x_{c}>=(a^{2}+w^{2}t^{2})%
\delta _{bc}$, where $\delta _{bc}$ is the unit tensor. So, we get 3-D
Gaussian distribution for the position of particle with indicated moments.

Tachyonic impulses can be at the core of the phenomena of quantum
entanglement, as indicated before [3]. Consider system of $n$ particles,
positions of which $\mathbf{x=(}x_{1},...,x_{n})$ are initially correlated.
We start again with 1-D case and 3-D generalization will follow. We assume
that particles do not interact directly. The first possibility is that
particles may not interact because of their nature. The second possibility
is that we separate them initially, so that probability of direct collision
is low (here 3-D generalization is handy). We choose initial condition such
that mean positions $<\mathbf{x}>$ will not change in time (see below). To
simplify notation, we assume that $<\mathbf{x}>$ is already extracted from $%
\mathbf{x}$. For the corresponding wave function $\Psi (t,\mathbf{x})$,
instead of (3), we now have equation:

\begin{equation}
\frac{\partial \Psi }{\partial t}=i\sum q_{j}\frac{\partial ^{2}\Psi }{%
\partial x_{j}^{2}},\;q_{j}\equiv \frac{\text{%
%TCIMACRO{\U{127}}%
%BeginExpansion
h{\hskip-.2em}\llap{\protect\rule[1.1ex]{.325em}{.1ex}}{\hskip.2em}%
%EndExpansion
}}{2m_{j}}  \tag{24}
\end{equation}

where summation is over $j=1,...,n$ and $m_{j}$ are masses of particles.
Fourier transformation of (24) gives:

\begin{equation}
\partial \widetilde{\Psi }/\partial t=-i[\sum q_{j}k_{j}^{2}]\widetilde{\Psi 
}  \tag{25}
\end{equation}

Here $\widetilde{\Psi }(t,\mathbf{k})$ is the transform of $\Psi (t,\mathbf{x%
})$ and $\mathbf{k}=(k_{1},...,k_{n})$ is the wave number vector. Solution
of (25) is:

\begin{equation}
\widetilde{\Psi }(t,\mathbf{k})=\widetilde{\Psi }_{0}(\mathbf{k})\exp
\{-it[\sum q_{j}k_{j}^{2}]\}  \tag{26}
\end{equation}

We assume initial Gaussian distribution for positions of particles [2]:%
\begin{equation}
P_{0}(\mathbf{x})=\frac{1}{(2\pi )^{n/2}(Det\mathbf{C})^{1/2}}\exp \{-\frac{1%
}{2}\mathbf{xC}^{-1}\mathbf{x}\}  \tag{27}
\end{equation}

Here $\mathbf{C=}C_{jl}=<x_{j}x_{l}>$ is the initial covariance matrix
(symmetric) and $\mathbf{C}^{-1}$is the inverse matrix. In particular, $%
C_{11}=<x_{1}^{2}>=a_{1}^{2}$ is dispersion for position of the first
particle with zero mean ($<x_{1}>=0$), $a_{j}$ corresponds to the particle $%
j $, $\rho _{jl}(0)=C_{jl}(a_{j}a_{l})^{-1}$ (with $j\neq l$) is the initial
correlation coefficient. Characteristic function for (27) is:

\begin{equation}
\Phi _{0}(\mathbf{k})=<\exp \{i\mathbf{kx}\}>=\exp \{-\frac{1}{2}\mathbf{kCk}%
\}  \tag{28}
\end{equation}

Choosing initial wave function $\Psi _{0}(\mathbf{x})=[P_{0}(\mathbf{x)]}%
^{1/2}$, after simple calculation, we get transform:

\begin{equation}
\widetilde{\Psi }_{0}(\mathbf{k})=2^{3n/4}\pi ^{n/4}(Det\mathbf{C}%
)^{1/4}\exp \{-\mathbf{kCk}\}  \tag{29}
\end{equation}

Formulas (26) and (29) give:%
\begin{equation}
\widetilde{\Psi }(t,\mathbf{k})=2^{3n/4}\pi ^{n/4}(Det\mathbf{C})^{1/4}\exp
\{-\mathbf{k}[\mathbf{C}+it\mathbf{qI]k}\}  \tag{30}
\end{equation}

Here $\mathbf{qI}$ stands for diagonal matrix with elements $q_{j}$. This
expression (apart from constant prefactor) is characteristic function for
the joint epd of $n$ complex-Gaussian trajectories like (10) with
statistically independent velocities $v_{j}(t)$. It means, in particular,
that $<v_{1}(t+\tau )v_{2}(t)>=<v_{1}(t+\tau )><v_{2}(t)>\equiv 0$ and
mutual correlation function of these trajectories is equal to correlation of
initial positions. Such simple is the quantum entanglement in the "complex
world". In the real world situation is more complicated, as we will see
below.

Consider characteristic function for $P(t,\mathbf{x})$ by using
generalization of (18):

\begin{equation}
\Phi (t,\mathbf{k})=\frac{1}{(2\pi )^{n}}\int d\mathbf{m}\widetilde{\Psi }(t,%
\mathbf{m})\widetilde{\Psi }^{\ast }(t,\mathbf{m-k})  \tag{31}
\end{equation}

Substitution of (30) into (31) gives:

\[
\Phi (t,\mathbf{k})=(\frac{2}{\pi })^{n/2}(Det\mathbf{C})^{1/2}\int d\mathbf{%
m}\exp \{-2\mathbf{mCm+}2\mathbf{mCk}-2it\mathbf{m\varkappa }+it\mathbf{%
k\varkappa }-\mathbf{kCk\}} 
\]%
where we used symmetry of $\mathbf{C}$ and introduced vector $\mathbf{%
\varkappa }=(q_{1}k_{1},...,q_{n}k_{n})$. By substitution $\mathbf{%
m\varkappa }=\mathbf{mCC}^{-1}\mathbf{\varkappa }$, the integral takes
standard Gaussian form:%
\[
\Phi (t,\mathbf{k})=(\frac{2}{\pi })^{n/2}(Det\mathbf{C})^{1/2}\int d\mathbf{%
\mu }\exp \{-2\mathbf{\mu C\mu }+\frac{1}{2}\mathbf{\chi C\chi }+it\mathbf{%
k\varkappa }-\mathbf{kCk\}} 
\]

where $\mathbf{\mu }=\mathbf{m}-\mathbf{\chi }/2,\;\mathbf{\chi }=\mathbf{k}%
-it\mathbf{C}^{-1}\mathbf{\varkappa }$. The linear time-dependence of
dispersions is eliminated by the interaction of the tachyonic impulses, as
can be anticipated. But, there are interesting details, which are hard to
predict without calculations. The integration gives characteristic function,
which corresponds to Gaussian distribution:

\begin{equation}
\Phi (t,\mathbf{k})=\exp \{-\frac{1}{2}[\mathbf{kCk}+t^{2}\mathbf{\varkappa C%
}^{-1}\mathbf{\varkappa }]\}  \tag{32}
\end{equation}

For $n=1$, formula (32) reproduces (23). For $n=2$ formula (32) gives:

\begin{equation}
\Phi (t,\mathbf{k})=\exp \{-\frac{1}{2}[\sigma _{1}^{2}(t)k_{1}^{2}+2\rho
(t)\sigma _{1}(t)\sigma _{2}(t)k_{1}k_{2}+\sigma _{2}^{2}(t)k_{2}^{2}]\} 
\tag{33}
\end{equation}

\begin{equation}
\sigma _{1}^{2}(t)=a_{1}^{2}+w_{1}^{2}t^{2},\;\sigma
_{2}^{2}(t)=a_{2}^{2}+w_{2}^{2}t^{2},\;w_{1}=q_{1}(a_{1}\beta
_{0})^{-1},\;w_{2}=q_{2}(a_{2}\beta _{0})^{-1}  \tag{34}
\end{equation}

\begin{equation}
\rho (t)=\rho _{0}\frac{a_{1}a_{2}-w_{1}w_{2}t^{2}}{\sigma _{1}(t)\sigma
_{2}(t)}  \tag{35}
\end{equation}

Here $\sigma _{1}^{2}$ and $\sigma _{2}^{2}$ are corresponding dispersions, $%
\rho $ is correlation coefficient with initial value $\rho _{0}$ and $\beta
_{0}=(1-\rho _{0}^{2})^{1/2}$. The first important feature of this
distribution is the factor $\beta _{0}$ in (34) - increase of velocities due
to initial correlation. In terms of vortices (see above), it could mean such
interaction that, in average, gets them closer to real line.

Additionally, we get evolution of correlation coefficient $\rho (t)$ from
initial value $\rho _{0}$ to asymptotic value $-\rho _{0}$ (when $t\gg \max
\{a_{1}/w_{1},a_{2}/w_{2}\}$). We can say that quantum vacuum "does not
like" the measurement-imposed correlation and makes the inversion.

It will be interesting to play with these features experimentally.
Particularly, it is important to investigate if the imposing of initial
correlation can be made energy-efficient. Then the shooting effect with high
velocities could be a potential mechanism for obtaining energy.

The 3-D generalization for the two-particle system is as follows. Positions
of particles and wave numbers we denote by $x_{j,b}$and $k_{j,b}$, where $%
j=(1,2)$ indicate particle and $b=(1,2,3)$ corresponds to 3-D vector. Matrix
of initial correlations now has two sets of indexes: $C_{jl,bc}$. In
simplest case: $C_{jl,bc}=C_{jl}\delta _{bc}$. In this case we can use (33)
with products $k_{1}^{2}$, $k_{1}k_{2}$ and $k_{2}^{2}$ replaced by the
scalar products of vectors, in particular, $k_{1}k_{2}$ is replaced by $%
k_{1,1}k_{2,1}+k_{1,2}k_{2,2}+k_{1,3}k_{2,3}$. From such generalization of
(33) it follows: $<x_{j,b}>=0,\;<x_{1,b}x_{1,c}>=\sigma _{1}^{2}\delta _{bc}$
$,\;<x_{2,b}x_{2,c}>=\sigma _{2}^{2}\delta _{bc}$, $<x_{1,b}x_{2,c}>=\rho
\sigma _{1}\sigma _{2}\delta _{bc}$.

In general case formula (32) gives:

\begin{equation}
<x_{j,b}x_{l,c}>=C_{jl,bc}+t^{2}q_{j}q_{l}C_{jl,bc}^{-1}  \tag{36}
\end{equation}

In particular:

\begin{equation}
<x_{1,1}^{2}>=C_{11,11}+t^{2}q_{1}^{2}C_{11,11}^{-1}  \tag{37}
\end{equation}

Velocity of the shooting in (37) is determined by the term of the inverse
covariance matrix, which is multiplied by $x_{1,1}^{2}$ in the initial
probability density (27). For the two-particle system this term, according
to (34), is proportional to $(1-\rho _{0}^{2})^{-1}$. For $n\geqslant 3$
velocity of the shooting depends of all initial correlation between
particles.

The presented results can be also obtained and interpreted in terms of the
Wigner function (see, for example, recent book [4] and references therein):

\begin{equation}
W(t,x,p)=\frac{1}{2\pi \text{%
%TCIMACRO{\U{127}}%
%BeginExpansion
h{\hskip-.2em}\llap{\protect\rule[1.1ex]{.325em}{.1ex}}{\hskip.2em}%
%EndExpansion
}}\int dr\Psi (t,x-r/2)\Psi ^{\ast }(t,x+r/2)\exp \{\frac{ipr}{\text{%
%TCIMACRO{\U{127}}%
%BeginExpansion
h{\hskip-.2em}\llap{\protect\rule[1.1ex]{.325em}{.1ex}}{\hskip.2em}%
%EndExpansion
}}\}  \tag{38}
\end{equation}

Here we consider the 1-D case for one particle and $p$ is additional
(momentum) variable. The probability density is given by integral:

\begin{equation}
\Psi \Psi ^{\ast }=\int dpW  \tag{39}
\end{equation}

Solution (5) [or (6)] corresponds to:

\begin{equation}
W(t,x,p)=W_{0}(x-\frac{pt}{m},p)  \tag{40}
\end{equation}

Initial condition (12) gives:

\begin{equation}
W_{0}(x,p)=\frac{1}{\pi \text{%
%TCIMACRO{\U{127}}%
%BeginExpansion
h{\hskip-.2em}\llap{\protect\rule[1.1ex]{.325em}{.1ex}}{\hskip.2em}%
%EndExpansion
}}\exp \{-\frac{x^{2}}{2a^{2}}-\frac{2a^{2}p^{2}}{\text{%
%TCIMACRO{\U{127}}%
%BeginExpansion
h{\hskip-.2em}\llap{\protect\rule[1.1ex]{.325em}{.1ex}}{\hskip.2em}%
%EndExpansion
}^{2}}\}  \tag{41}
\end{equation}

Using (40), we have:

\begin{equation}
W(t,x,p)=\frac{1}{\pi \text{%
%TCIMACRO{\U{127}}%
%BeginExpansion
h{\hskip-.2em}\llap{\protect\rule[1.1ex]{.325em}{.1ex}}{\hskip.2em}%
%EndExpansion
}}\exp \{-\frac{(x-ptm^{-1})^{2}}{2a^{2}}-\frac{2a^{2}p^{2}}{\text{%
%TCIMACRO{\U{127}}%
%BeginExpansion
h{\hskip-.2em}\llap{\protect\rule[1.1ex]{.325em}{.1ex}}{\hskip.2em}%
%EndExpansion
}^{2}}\}  \tag{42}
\end{equation}

Integration (39) of (42) reproduces (15). In addition, we get interpretation
of the random shooting (16) in terms of variable $p$. Let us note that
formula (15) for one particle was known before (see references in [4]), but
interpretation was different. Expression $(a^{2}+w^{2}t^{2})^{1/2}$ (in our
notation) was interpreted as trajectory [4]. In our interpretation, the
motion is with constant (random) velocity (16).

The $n$-particle system has analogous description by the Wigner function
with interpretation of general formulas (32) and (36) in terms of
corresponding $p$-variables. It is helpful to have in mind both approaches (
Schr\"{o}dinger and Wigner ) in order to develop intuition, which unites
tachyonic impulses and $p$-variables.

We plan detailed investigation of systems with three or more particles by
using presented above general results.

Further studies of ID can also open new perspective in quantum field theory.

I thank V. I. Tatarskii for useful discussion.

\bigskip

\bigskip

\bigskip

\textbf{References}

\bigskip

[1] L.D. Landau and E. M. Lifshitz, Quantum Mechanics, Pergamon Press, 1987

[2] N. G. Van Kampen, Stochastic Processes in Physics and Chemistry,
North-Holland, 1985

[3] E. A. Novikov, Vacuum response to cosmic stretching: accelerated
universe and prevention of singularity, arXiv:nlin.PS/0608050

[4] R. E. Wyatt, Quantum Dynamics with Trajectories, Springer, 2005

\end{document}